\documentclass[floatfix,twocolumn,showpacs,amsmath,amssymb]{revtex4}
\usepackage{graphicx}
\usepackage{times}

\DeclareMathOperator{\Sch}{S}  % Schwarzian derivative

\newcommand{\fwidth}{.45\textwidth}  % Figure width for two-column

\begin{document} 

\title{Robust chaos with variable Lyapunov exponent in smooth one-dimensional maps}
\author{Juan~M.~Aguirregabiria, \email{juanmari.aguirregabiria@ehu.es}} 
\affiliation{Theoretical Physics, 
The University of the Basque Country, \\
P.~O.~Box 644,
48080 Bilbao, Spain}
\email{juanmari.aguirregabiria@ehu.es} 

\date{\today}% It is always \today, today,

%%%%%%%%%%%%%%%%%%%%%%%%%%%%%%%%%%%%%%%%%%%%%%
\begin{abstract} 
We present several new easy ways of generating smooth one-dimensional maps displaying robust chaos,
i.e., chaos for whole intervals of the parameter. Unlike what happens with previous methods,
the Lyapunov exponent of the maps constructed here varies widely with the parameter.
We show that the condition of 
negative Schwarzian derivative, which was used in previous works,
is not a necessary condition for robust chaos. Finally we show that the maps constructed
in previous works have always the Lyapunov exponent $\ln 2$ because they 
are conjugated to each other and to the tent map by means of smooth
homeomorphisms. In the methods presented here, the maps have variable Lyapunov coefficients because
they are conjugated
 through non-smooth homeomorphisms similar to
Minkowski's question mark function.
\end{abstract} 



\maketitle

%%%%%%%%%%%%%%%%%%%%%%%%%%%%%%%%%%%%%%%%%%%%%%
\section{\label{sec:intro}Introduction}
%%%%%%%%%%%%%%%%%%%%%%%%%%%%%%%%%%%%%%%%%%%%%%

Many families of smooth maps display fragile chaos, which may be
destroyed by arbitrarily small changes of the parameter. For instance,
the discrete dynamical system generated by the logistic map,
$x_{n+1}=\mu x_n\left(1-x_n\right)$, is chaotic for $\mu=4$, but the
values of the parameter $\mu$ for which the attractor is periodic are
dense in the interval $[0,4]$ \cite{Ott}. In consequence, when such a
family is used to describe a physical system, it may be impossible to
decide on theoretical grounds whether the actual behavior of the system
will be chaotic or periodic for some parameter value, which is necessarily known
only approximately. Furthermore, some practical applications, 
such as encoding messages \cite{Hayes}, require reliable chaotic
behavior. 

Piecewise smooth maps may show robust chaos and
they have been used to describe a circuit with robust chaotic output \cite{Banerjee}.
Although for some time it was conjectured that one-dimensional maps should be piecewise
smooth to display robust chaos \cite{Barreto,Banerjee}, Andrecut and Ali first found a smooth map
\cite{Andrecut1} and later a method of generating smooth maps \cite{Andrecut2} whose  evolution is
chaotic for whole intervals of the parameter.

The purpose of this work is twofold: we want to explore other easy ways of generating robust chaos in one-dimensional
smooth maps and to check whether the condition of negative Schwarzian derivative satisfied by the maps
of Refs.\ \cite{Andrecut1,Andrecut2} is a necessary one. Unlike in previous methods, the
Lyapunov exponent of the maps explored here takes rather different values depending on the value of the parameter. This property
might be an advantage in some applications. We will see in Sect.\ \ref{sec:conjugate} 
that the reason of this dependence lies 
in the different ways in which maps generated by each method are conjugate
to each other.

We will consider maps on a finite interval $[a,b]$, which for commodity will be reduced to $[0,1]$
 by means of a linear transformation.

%%%%%%%%%%%%%%%%%%%%%%%%%%%%%%%%%%%%%%%%%%%%%%
\section{\label{sec:schw}Robust chaos with negative Schwarzian derivative}
%%%%%%%%%%%%%%%%%%%%%%%%%%%%%%%%%%%%%%%%%%%%%%

The Schwarzian derivative of the function $f$ is defined as
\begin{equation}
\Sch f(x)\equiv\frac{f'''(x)}{f'(x)}-\frac32\left(\frac{f''(x)}{f'(x)}\right)^2.
\end{equation}
Since Singer used it in the study of the bifurcations of maps of the interval \cite{Singer},
a key assumption in many theorems on the dynamics of one-dimensional discrete dynamical systems
 is that the Schwarzian derivative of the map is negative
along the whole interval. 

In the following we will take advantage of the fact that the Schwarzian derivative
is invariant under linear fractional transformations \cite{Jackson}, i.e., that for 
constants $a$, $b$, $c$ and $d$ one has
\begin{equation}\label{eq:invariance}
 \Sch\frac{af(x)+b}{cf(x)+d}=\Sch f(x).
\end{equation}

Our starting point will be a map $f: [0,1]\to[0,1]$ of class $C^3$, which we assume to be
`S-unimodal,' i.e.,  which satisfies $f(0)=f(1)=0$,
 has a single critical point at $c\in(0,1)$, and negative Schwarzian derivative everywhere in $[0,1]$.
Clearly $f$ increases from its null value at $x=0$ until it reaches its maximum at $x=c$,
and then decreases until becoming 0 again at $x=1$.
According to Singer's theorem \cite{Singer,Jackson}, the discrete dynamical system
$x_{n+1}=f(x_n)$ has at most one stable periodic orbit, which when exists attracts
the critical point $x=c$. 
%In the following we shall consider as examples for $f$
%the `logistic' map $f(x)=\mu x(1-x)$ and the `asymmetric' map $f(x)=\mu x\left(1-x^2\right)$.

For any map $f$ with the properties above mentioned we will construct the following one-parameter family of maps:
\begin{equation}\label{eq:family1}
f_r(x)\equiv \frac{(1+r)f(x)}{f(c)+rf(x)},\qquad (-1<r<\infty).
\end{equation}
(Notice that $f_r$ does not change if one multiplies $f$ by any constant.)

By using (\ref{eq:invariance}) and
\begin{equation}
 f_r'(x)= \frac{(1+r)f(c)f'(x)}{[f(c)+rf(x)]^2}
\end{equation}
one can readily check that $f_r$ is also S-unimodal for all $r>-1$.

Now, for 
\begin{equation}\label{eq:conditionst}
r>r_0\equiv\frac{f(c)}{f'(0)}-1,
\end{equation}
 the origin is a unstable fixed point, because
then $f'_r(0)>1$ and, furthermore, the dynamical system $x_{n+1}=f_r(x_n)$ has no
stable periodic orbit, because the critical point goes, in two steps, to the unstable
origin: $f^2_r(c)=0$.  
%%%%%%%%%%%%%%%%%%%%%%%%%%%%%%%%%%%%%%%%%%%%%
% Figure 1
\noindent\begin{figure}
\begin{center}
\noindent\includegraphics[width=\fwidth]{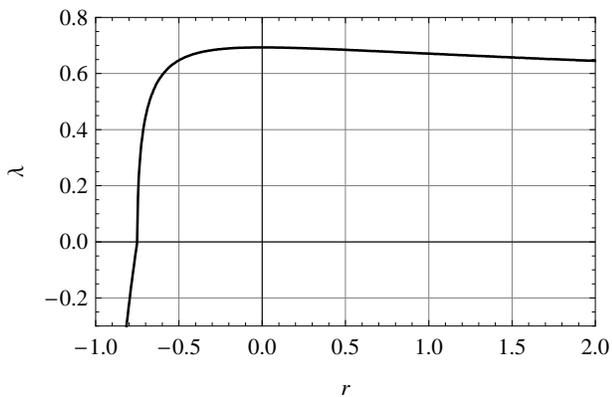}
\end{center}
\caption{Lyapunov exponent of the map (\ref{eq:family1}) for $f(x)=x(1-x)$.\label{fig1}} 
\end{figure}
%%%%%%%%%%%%%%%%%%%%%%%%%%%%%%%%%%%%%%%%%%%%%

To check that the dynamical system is chaotic for all $r>r_0$ on can compute numerically 
 the Lyapunov exponent
\begin{equation}
\lambda = \lim_{N\to\infty}\sum_{n=1}^N{\ln\left|f_r'\left(x_n\right)\right|}.
\end{equation}
With the maps $f_r$ generated from the logistic map $f(x)=x(1-x)$ by means of (\ref{eq:family1}), one gets the values displayed in Fig.\ \ref{fig1}. They
are negative for $r<r_0=-3/4$, because then the origin is an attractor, and become positive at $r>r_0$ as
the generic orbit wanders chaotically around the whole interval. In the maps generated in
Refs.\ \cite{Andrecut1,Andrecut2} the Lyapunov coefficient was always $\ln 2$ or very close,
according to the numerical simulations. In the maps presented above, the Lyapunov exponent varies with $r$
in a continuous way. If $f(x)=x(1-x)$ the maximum value of the Lyapunov exponent of $f_r(x)$ is
 $\lambda_\mathrm{max} =\ln 2$. This value is reached at $r=0$, which 
corresponds to the well known case $x_{n+1}=4x_n\left(1-x_n\right)$, which in turn is conjugate
to the tent map defined as $f(x)=2x$ for $0\le x\le1/2$
and $f(x)=2-2x$ for $1/2\le x\le 1$ \cite{Ott}. 
Similar graphs, with $0<\lambda\le\lambda_\mathrm{max} =\ln 2$, 
are obtained, for instance, 
for the asymmetric map $f(x)=x(1-x^2)$ and for $f(x)=\sin \pi x$, although the bifurcation
value $r_0$ and the location of the maximum 
are different. 
%%%%%%%%%%%%%%%%%%%%%%%%%%%%%%%%%%%%%%%%%%%%%
% Figure 2
\begin{figure}
\begin{center}
\includegraphics[width=\fwidth]{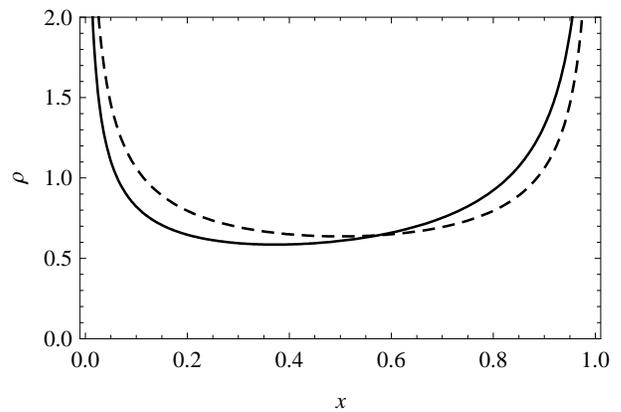}
\end{center}
\caption{Natural invariant density of the map (\ref{eq:family1}) for $f(x)=x(1-x)$ and $r=0$ (dashed line)
and $r=1$ (continuous line).\label{fig2}} 
\end{figure}
%%%%%%%%%%%%%%%%%%%%%%%%%%%%%%%%%%%%%%%%%%%%%

One can also compute numerically the natural invariant measure $d\mu=\rho_r(x)\,dx$ by using the 
Frobenius-Perron equation \cite{Ott}
satisfied by the
natural invariant density $\rho_r(x)$:
\begin{equation}
%\rho_r(x)=\sum_{i=1}^2{\frac{\rho_r\left(y_i\right)}{\left|f'_r\left(y_i\right)\right|}},
\rho_r(x)=\frac{\rho_r\left(y_1\right)}{\left|f'_r\left(y_1\right)\right|}+\frac{\rho_r\left(y_2\right)}{\left|f'_r\left(y_2\right)\right|},
\end{equation}
where $y_1$ and $y_2$ are the preimages of $x$, i.e., $f_r\left(y_1\right)=f_r\left(y_2\right)=x$.
In the case of  $f(x)=x(1-x)$ it is well known that for the logistic map $f_0$ the natural invariant density is
$\rho_0(x)=\left[\pi^2x(1-x)\right]^{-1/2}$. It is displayed
in Fig.\ \ref{fig2}, along with the natural invariant density for $f_1(x)$. Only in the
first case (for $r=0$) is the natural invariant density symmetric around the critical point $x=1/2$.
Similar results are obtained with other choices of $f(x)$.

When constructing smooth maps by using
the method of Andrecut and Ali \cite{Andrecut2} or the one provided by Eq.\ (\ref{eq:family1}),
robust chaos is guaranteed by Singer's theorem; but they are by no means the only 
way to get chaos for an interval of the parameter. 
For instance, we have been exploring the family generated from
a S-unimodal map $f(x)$ by the expression
\begin{equation}\label{eq:family2}
f_r(x)\equiv\frac{1+r(x-c)^2}{f(c)}\,f(x).
\end{equation}
The Schwarzian derivative $\Sch f_r$ has a rather involved expression which makes difficult, if not impossible,
a general analysis. However, selecting the logistic map $f(x)=x(1-x)$, 
it is easy to see that the corresponding $f_r(x)$ is S-unimodal for $-4<r<4$ and displays
robust chaos for $-3<r<4$. The plot of the corresponding Lyapunov exponent is very
similar to that of Fig.\ \ref{fig1} (including the location and the value of its maximum), except
for the fact that the bifurcation happens at $r=-3$. 
Similar results have been obtained with
$f(x)=x\left(1-x^2\right)$ and $f(x)=\sin\pi x$.

%%%%%%%%%%%%%%%%%%%%%%%%%%%%%%%%%%%%%%%%%%%%%%
\section{\label{sec:other}Robust chaos with positive Schwarzian derivative}
%%%%%%%%%%%%%%%%%%%%%%%%%%%%%%%%%%%%%%%%%%%%%%

All the maps discussed above, as well as those of Andrecut and Ali \cite{Andrecut1,Andrecut2}
and the `B-Exponential' map of Ref.\ \cite{Shastry},
satisfy the condition of negative Schwarzian derivative. This is a very powerful condition, 
but also rather restrictive and can be destroyed by a smooth change of the $x$ coordinate \cite{Melo}
or by small perturbations \cite{Jackson}.
In consequence, it may be of practical interest to find robust chaos in
one-dimensional maps even when that condition is not satisfied. We will see in the following
that the condition is not necessary to have robust chaos in one-dimensional smooth maps.

%%%%%%%%%%%%%%%%%%%%%%%%%%%%%%%%%%%%%%%%%%%%%
% Figure 3
\begin{figure}
\begin{center}
\includegraphics[width=\fwidth]{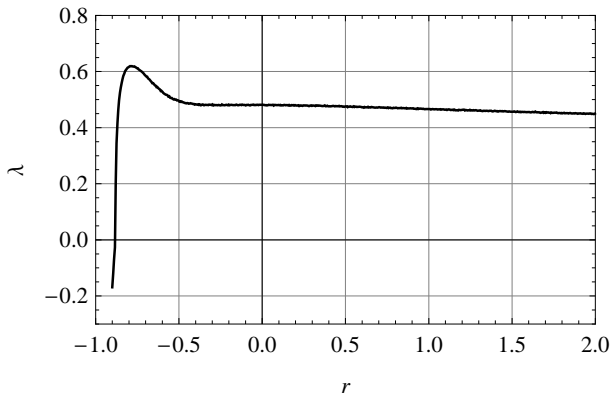}
\end{center}
\caption{Lyapunov exponent of the map (\ref{eq:family1}) for Singer's function (\ref{eq:singer}).\label{fig3}} 
\end{figure}
%%%%%%%%%%%%%%%%%%%%%%%%%%%%%%%%%%%%%%%%%%%%%
Let us first consider Singer's function \cite{Jackson}
\begin{equation}\label{eq:singer}
f(x) = 7.86 x-23.31x^2+28.75x^3-13.3x^4,
\end{equation}
and the map $f_r$ generated from it by means of (\ref{eq:family1}).
Since $f(x)$ has a positive Schwarzian derivative
in a subinterval of $[0,1]$, exactly the same happens with $f_r(x)$. 
However, a numerical computation of the Lyapunov exponent
shows that $f_r(x)$ has robust chaos after the origin becomes unstable at $r\approx-0.88156$.
In Fig.\ \ref{fig3} we can see that the 
maximum value of the Lyapunov coefficient is in this case $\lambda_\mathrm{max}\approx 0.62$,
i.e., somewhat smaller than the maximum value $\ln 2$ obtained in all previous examples.
As happens in those examples,  there is no attractor as the generic orbit wanders chaotically around
the whole interval.

%%%%%%%%%%%%%%%%%%%%%%%%%%%%%%%%%%%%%%%%%%%%%
% Figure 4
\begin{figure}
\begin{center}
\includegraphics[width=\fwidth]{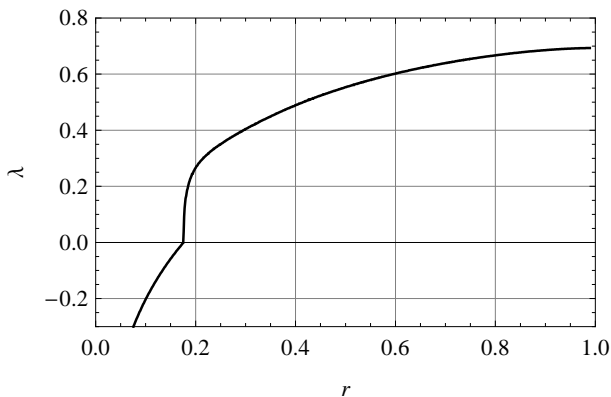}
\end{center}
\caption{Lyapunov exponent of the map (\ref{eq:family3}) for the logistic map.\label{fig4}} 
\end{figure}
%%%%%%%%%%%%%%%%%%%%%%%%%%%%%%%%%%%%%%%%%%%%%
We have also explored the one-parameter family of maps
\begin{equation}\label{eq:family3}
f_r(x)\equiv \left(\frac{f(x)}{f(c)}\right)^r,\qquad(r>0),
\end{equation}
for some choices of $f(x)$. 

In the case of the logistic map $f(x)=x(1-x)$, the map $f_r(x)$ is S-unimodal
only when $r=1$. 
For $r>1$ the function has a minimum at the origin. In consequence, $x=0$ is a stable fixed point that
attracts the generic orbit, after a chaotic transient, which may be very long for values of $r$ just above 1, for then the
basin of attraction of $x=0$ is tiny.
For $0<r<1$ the Schwarzian derivative is positive near the origin and $x=1$. For instance,
\begin{equation}
\Sch f_r(x)\sim \frac{1-r^2}{2x^2}\quad\mbox{as } x\to0.
\end{equation}
Furthermore, the map is not even $C^3$ in that case, because its first derivative goes to infinity at $x=0,\ 1$.
However, we can see in Fig.\ \ref{fig4} that robust chaos arises after the fixed point located in the interval $(1/2,1)$
becomes unstable at $r\approx0.1759$. The maximum Lyapunov exponent is again $\ln 2$ and is reached at
$r=1$, when we recover the logistic map $f(x)=4x(1-x)$. Similar results have been obtained with
$f(x)=x\left(1-x^2\right)$ and $f(x)=\sin\pi x$. 

The examples discussed in this section suggest 
that robust chaos may not be an unusual property of smooth one-dimensional maps, even when the condition
of negative Schwarzian derivative is not satisfied.  
 
%%%%%%%%%%%%%%%%%%%%%%%%%%%%%%%%%%%%%%%%%%%%%%
\section{\label{sec:conjugate}Robust chaos and conjugate maps}
%%%%%%%%%%%%%%%%%%%%%%%%%%%%%%%%%%%%%%%%%%%%%%

All the maps generated here and in previous works have qualitatively similar dynamics:
the solution wanders around the whole interval in a chaotic way. Furthermore, the graphs
of all the maps $g=f_r$ look rather similar: they start from $g(0)=0$, increase monotonically
until $g(c)=1$ and the decrease monotonically until $g(1)=0$. This suggest all the 
maps are conjugate \cite{Ott} to each other, i.e., given two of these maps, $g$ and $\tilde g$, 
there exist a homeomorphism $\phi$ on $[0,1]$ such that $\tilde g=\phi\circ g\circ\phi^{-1}$.
In other words, 
there exists a continuous change of variables $x\to\tilde x=\phi(x)$, with continuous inverse,
such that
\begin{equation}\label{eq:conjugate}
\phi\left[g(x)\right]=\tilde g\left[\phi(x)\right],\quad \forall x\in[0,1].
\end{equation}
The dynamical systems $x_{n+1} =g\left(x_n\right)$ and  $\tilde x_{n+1} =\tilde g\left(x_n\right)$
have essentially equivalent dynamics (for instance, if the unstable periodic orbits are dense for
$g$ the same will happen for $\tilde g$).
If, additionally, $\phi$ and $\phi^{-1}$ are smooth both maps have
the same Lyapunov coefficient \cite{Ott}.

This property provides a simple method of constructing families of maps with robust chaos and constant Lyapunov exponent:
take a chaotic map $f(x)$ and a smooth homeomorphism $\phi(x)$ on $[0,1]$ depending continuously on a 
parameter $r$. Them $f_r\equiv \phi\circ f\phi^{-1}$ will have the same Lyapunov exponent for all values of $r$.
For instance, using $f(x)=4x(1-x)$ and $\phi(x)=x^r$ for $r>0$ we get
\begin{equation}\label{eq:familyc}
f_r(x)=4^rx\left(1-x^{1/r}\right)^r,\qquad (r>0),
\end{equation}
whose Lyapunov exponent will be $\lambda=\ln2$ for all $r>0$. However, this is not S-unimodal, except for $r=1$, because
its Schwarzian derivative becomes positive near $x=0$ for $r>1$ and near $x=1$ for $0<r<1$.  
(Moreover, $f_r(1)>0$ for $0<r<1$.)
We see again that
the condition of negative Schwarzian derivative is not conserved by smooth changes of coordinates
and is not necessary for robust chaos. Notice that,  by using the well known results 
corresponding to the full logistic map \cite{Ott}, we can  write explicitly
the solution of the dynamical system driven by the map (\ref{eq:familyc}) 
as  $x_n=\sin^{2r}\left(2^n\arcsin x_0^{1/2r}\right)$ and  its natural invariant density as
$\rho(x)=\left[\pi^2r^2x^2\left(x^{-1/r}-1\right)\right]^{-1/2}$.
The `B-Exponential' map of Ref.\ \cite{Shastry} also is conjugate to the logistic map $g(x)=4x(1-x)$.

On the other hand, given two maps, $g$ and $\tilde g$, with the  qualitative properties mentioned at the 
start of this section, 
one can use a method of successive approximations
to construct the function $\phi$, if it exists. One may proceed as follows:
\begin{eqnarray}
\phi_0(x)&=&x,\label{eq:succesive1}\\
\phi_{n+1}(x)&=&\tilde g^{-1}\left[\phi_n\left(g(x)\right)\right],\quad (n=0,1,2,\ldots).\label{eq:succesive2}
\end{eqnarray}
The inverse function $\tilde g^{-1}$ is two-valued in this kind of map, but the right preimage is 
given by the condition that if  $c$ and $\tilde c$ are the
critical points of  $g$ and $\tilde g$ then $\phi(x) > \tilde c$ when $x>c$.
The method can be checked by computing $\phi^{-1}$ in the same way
to make sure that $\tilde g$ and $\phi\circ g\circ \phi^{-1}$ agree to the desired accuracy.

%%%%%%%%%%%%%%%%%%%%%%%%%%%%%%%%%%%%%%%%%%%%%
% Figure 5
\begin{figure}
\begin{center}
\includegraphics[width=\fwidth]{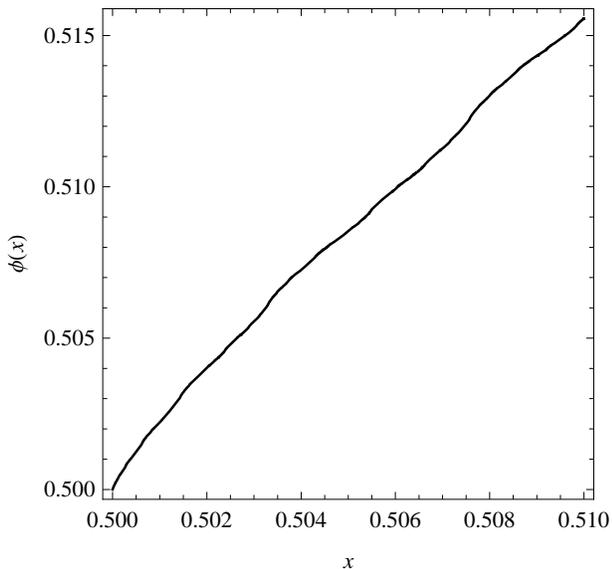}
\end{center}
\caption{Change of variables $\tilde x=\phi(x)$ when $\tilde g=4x(1-x)$ and $g$ is (\ref{eq:andrecut}) with 
$r=10$.\label{fig5}} 
\end{figure}
%%%%%%%%%%%%%%%%%%%%%%%%%%%%%%%%%%%%%%%%%%%%%

We have found that the method converges quickly when  $\tilde g(x) =4x(1-x)$ and $g$ is one of 
the maps generated in Refs.\ \cite{Andrecut1,Andrecut2}.
For instance, if we choose $g(x)$ as given by the map
\begin{equation}\label{eq:andrecut}
f_r(x)=\frac{1-r^{-x(1-x)}}{1-r^{-1/4}}
\end{equation}
of Ref.\ \cite{Andrecut2}, with $r=10$ and $\tilde g(x)=4x(1-x)$, we get the result of Figure \ref{fig5}.
Similar results are obtained with other values of $r$ and for the map in Ref.\ \cite{Andrecut1}.  
All these maps are thus conjugate to $f(x)=4x(1-x)$ and, in consequence \cite{Ott}, to the tent
map. Since the function $\phi$ and its inverse are smooth, all these maps share the Lyapunov exponent $\lambda =\ln2$.
%%%%%%%%%%%%%%%%%%%%%%%%%%%%%%%%%%%%%%%%%%%%%
% Figure 6
\begin{figure}
\begin{center}
\includegraphics[width=\fwidth]{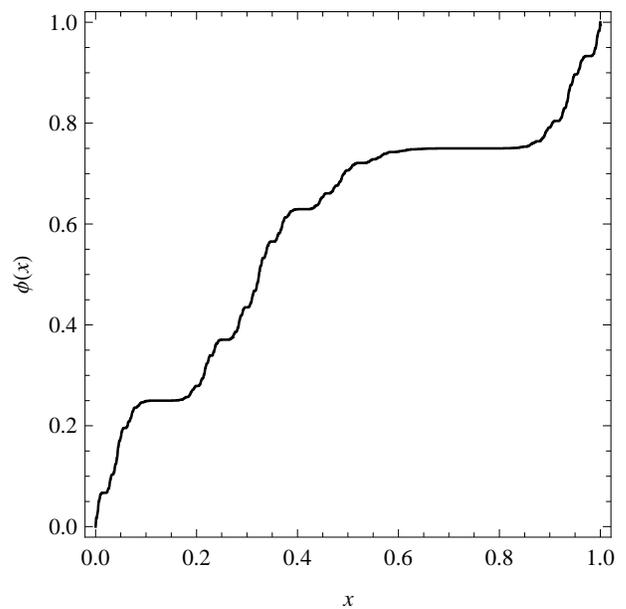}
\end{center}
\caption{Change of variables $\tilde x=\phi(x)$ when $\tilde g=4x(1-x)$ and $g$ is (\ref{eq:family1}), with $r=0$ and
$f(x)$ is given by (\ref{eq:singer}).\label{fig6}} 
\end{figure}
%%%%%%%%%%%%%%%%%%%%%%%%%%%%%%%%%%%%%%%%%%%%%

We have checked numerically that also the maps generated from $f(x)=x(1-x)$ by means of the different methods 
presented 
in this work are conjugate to $f(x)=4x(1-x)$ and, thus, to the tent map. But there is a crucial difference:
although $\phi$ and $\phi^{-1}$ are continuous they are not smooth enough for the two maps to share the
same Lyapunov exponent. For instance, in Fig.\ \ref{fig6} we have chosen 
$\tilde g=4x(1-x)$ and $g$ as given by (\ref{eq:family1}), with $r=0$, for Singer's map (\ref{eq:singer}).
It is clear there that $\phi'(x)$ vanishes at some points and, since the graph of $\phi^{-1}$ is 
obtained by exchanging the axes
of Fig.\ \ref{fig6}, the derivative of $\phi^{-1}(x)$ is infinite at those points. 

This explains why the corresponding Lyapunov exponents are different.
From this point of view, one can understand the methods of previous works as an easy way
to construct one-parameter families of conjugate maps by means of smooth homeomorphisms that guarantee
the conservation of the Lyapunov exponent, while the methods discussed here are easy ways
of constructing conjugate maps with Lyapunov exponents varying in a continuous way,
since they are conjugated by non-smooth homeomorphisms.

%%%%%%%%%%%%%%%%%%%%%%%%%%%%%%%%%%%%%%%%%%%%%
% Figure 7
\begin{figure}
\begin{center}
\includegraphics[width=\fwidth]{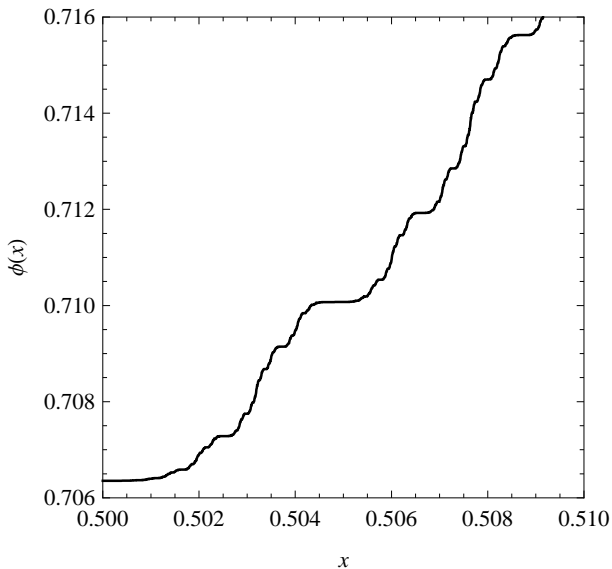}
\end{center}
\caption{Zoom of a $0.01\times0.01$ square of Fig.\ \ref{fig6}.\label{fig7}} 
\end{figure}
%%%%%%%%%%%%%%%%%%%%%%%%%%%%%%%%%%%%%%%%%%%%%

The complex structure of the graph of $\phi(x)$ in Fig.\ \ref{fig6} 
can be explored by zooming in on small parts of it. For instance, in Fig.\ \ref{fig7}
one can seen the function in the interval $[0.5,0.51]$. (A similar zoom of Fig.\ \ref{fig5}
reveals a smooth structure.)
 
\begin{figure}
\begin{center}
\includegraphics[width=\fwidth]{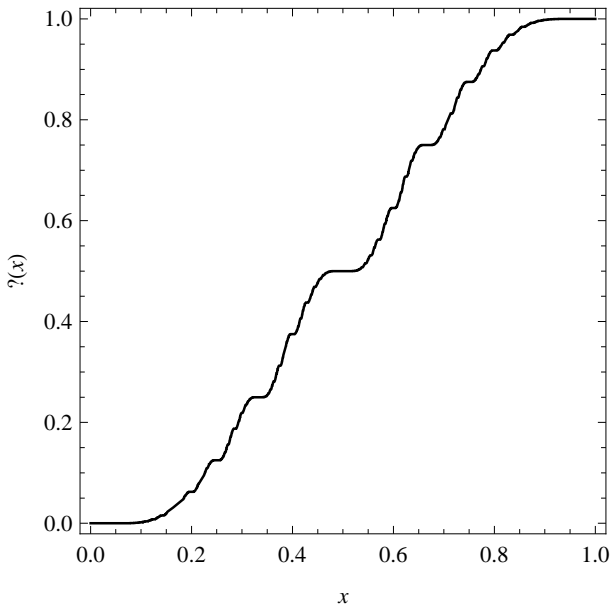}
\end{center}
\caption{Minkowski's question mark function.\label{fig8}} 
\end{figure}

In fact, the graph in Fig.\ \ref{fig6}, as well as the remaining graphs we have obtained with the maps 
generated by the methods presented in this work, looks very similar to the graph of Minkowski's question mark 
function  \cite{Minkowski} shown in Fig.\ \ref{fig8}. 
The resemblance is even more striking when both $g$ and $\tilde g$ are symmetric around the point $x=1/2$.

%%%%%%%%%%%%%%%%%%%%%%%%%%%%%%%%%%%%%%%%%%%%%
% Figure 8
%%%%%%%%%%%%%%%%%%%%%%%%%%%%%%%%%%%%%%%%%%%%%

Minkowski's  $?(x)$ function is continuous
with continuous inverse, strictly increasing and its derivative is zero almost everywhere and infinite
or undefined otherwise \cite{Viader,Paradis}. 
The question mark function is the homeomorphism $\phi$ conjugating the tent map and the Farey map \cite{Panti}
defined as $g(x)=x/(1-x)$ for $0\le x\le1/2$ and  $g(x)=(1-x)/x$ for $1/2\le x\le1$.
We have used this fact to check the accuracy of the method of successive approximations
given by (\ref{eq:succesive1})--(\ref{eq:succesive2}).
The fact that  $?(x)$ is not smooth explains the different Lyapunov exponents of the tent
map ($\tilde \lambda =\ln 2$) and the Farey map ($\lambda =0$). It also explains why the natural
invariant density of the latter map is not normalizable: $\rho(x)\propto x^{-1}$.

The numerical evidence we have found and the dependence on the parameter of the Lyapunov coefficient
strongly support the conjecture that the functions $\phi$ conjugating pairs of maps 
generated by the methods described in this work are not differentiable at an infinite number of points, 
probably almost everywhere.   

On the other hand, the fact that Minkowski's question mark function can be recursively
constructed by using the Farey sequence and continuity \cite{Girgensohn} suggests an alternative method
to construct $\phi$ for functions $g$ and $\tilde g$.
One starts from the critical point $x_0=c$, since we know $y_0\equiv\phi(x_0)=\phi(c)=\tilde c$. Then
for each pair $(x_n,y_n\equiv\phi(x_n))$ already computed, one can calculate two new pairs
\begin{equation}\label{eq:recr}
\left(x_{n+1},\ y_{n+1}\equiv\phi\left(x_{n+1}\right)\right)=
\left(g^{-1}_\pm\left(x_n\right),\ \tilde g^{-1}_\pm\left(y_n\right)
\right),
\end{equation}
where $g^{-1}_-(x)$ is the value $y$ satisfying  $g(y)=x$ and $y\le c$, 
while $y=g^{-1}_+(x)$ is given by the conditions $g(y)=x$ and $y> c$. Analogous definitions
are used for $\tilde g^{-1}_\pm$. We have checked that applying recursively (\ref{eq:recr}) one
 obtains again Figs.\ \ref{fig6} and~\ref{fig7}. The method also works 
for other pairs of maps constructed by means of (\ref{eq:family1}), (\ref{eq:family2}) or (\ref{eq:family3}).

%%%%%%%%%%%%%%%%%%%%%%%%%%%%%%%%%%%%%%%%%%%%%%
\section{\label{sec:final}Final comments}
%%%%%%%%%%%%%%%%%%%%%%%%%%%%%%%%%%%%%%%%%%%%%%

In previous examples ---including those of Refs.\ \cite{Andrecut1,Andrecut2} but
excluding (\ref{eq:familyc})--- the maximum
is located at the same point for all values of $r$; but this is not a necessary condition.
Let us consider the one-parameter family of maps
\begin{equation}\label{eq:family4}
f_r(x)\equiv\frac{f\left(x^r\right)}{f(c)},\qquad (0<r\le 1),
\end{equation}
which is obtained from family (\ref{eq:family3}) by means of the smooth homeomorphism
$\phi(x)=x ^{1/r}$. For instance,
if $f(x)=x(1-x)$, the maximum of (\ref{eq:family4}) is located at $x=\phi(1/2)=2^{-1/r}$ and the Lyapunov
exponent is that of Fig.~\ref{fig4} and its maximum value $\ln 2$ is reached again when $r=1$ and we recover
the full logistic map $f_1(x)=4x(1-x)$. 

%%%%%%%%%%%%%%%%%%%%%%%%%%%%%%%%%%%%%%%%%%%%%
% Figure 9
\begin{figure}
\begin{center}
\includegraphics[width=\fwidth]{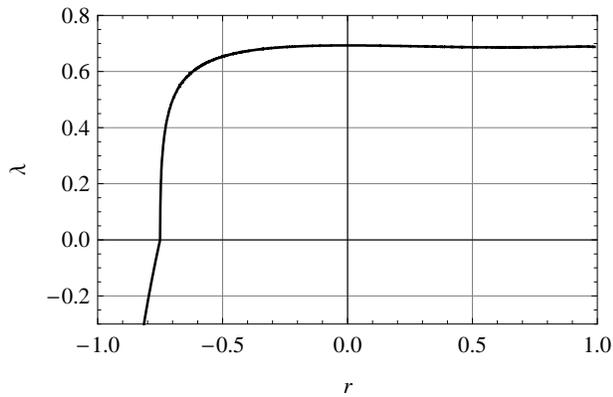}
\end{center}
\caption{Lyapunov exponent of the map (\ref{eq:family5}) for the logistic map.\label{fig9}} 
\end{figure}
%%%%%%%%%%%%%%%%%%%%%%%%%%%%%%%%%%%%%%%%%%%%%

We have also considered the following family of maps:
\begin{equation}\label{eq:family5}
f_r(x)\equiv \frac{f\left[(1+r)x-rx^2\right]}{f(c)},\qquad (-1\le r\le 1).
\end{equation}
If $f(x)=x(1-x)$, the maximum is located at $x=\left(\sqrt{1+r^2}-1+r\right)/(2r)$ and 
the origin becomes unstable for $r=-3/4$. Again the maximum Lyapunov coefficient 
is $\ln 2$, but it remains very close to this value for a large
parameter interval, as shown in Fig.~\ref{fig9}.

In all the examples considered above, as well as in those of Refs.\ \cite{Andrecut1,Andrecut2},
the Lyapunov exponent is never higher than $\ln2$; but it is easy to get other maximum
values by changing the starting map $f(x)$. Let consider only a simple example.
The piecewise linear map
\begin{equation}\label{eq:piece}
g(x)\equiv
\begin{cases}3x,&\mbox{if }0\le x\le 1/3;\\
                 2-3x;&\mbox{if }1/3\le x\le 2/3;\\
                 3x-2,&\mbox{if }2/3\le x\le 1
\end{cases}
\end{equation}
has $\left|g'(x)\right|=3$, except at $x=1/3,\ 2/3$. In consequence,
its Lyapunov exponent is $\ln 3$ and its natural invariant density $\rho(x)=1$.
If we use the change of variables $\tilde x=\phi(x)\equiv\sin^2(\pi x/2)$, 
the conjugate map $f\equiv \phi\circ g\circ\phi^{-1}$ is
\begin{equation}\label{eq:fpiece}
f(x)=x(4x-3)^2.
\end{equation}
The Lyapunov exponent of $f$ is $\ln 3$, because $\phi$ is smooth.
Since $\phi$ is precisely the map conjugating the tent map and the full logistic map $g(x)=4x(1-x)$,
the natural invariant density of the later is also that of $f$.
%%%%%%%%%%%%%%%%%%%%%%%%%%%%%%%%%%%%%%%%%%%%%%%
%% Figure 9
%%\begin{figure}
%%\begin{center}
%%\includegraphics[width=\fwidth]{fig9.eps}
%%\end{center}
%%\caption{Graph of the map $f(x)=x(4x-3)^2$.\label{fig9}} 
%%\end{figure}
%%%%%%%%%%%%%%%%%%%%%%%%%%%%%%%%%%%%%%%%%%%%%%

Although this map is qualitatively different from those considered above (for instance, it has two
critical points), the Lyapunov coefficient of the corresponding  family (\ref{eq:family1})
looks much like that in Fig.\ \ref{fig1}, except for the fact that the origin becomes
unstable at $r=-8/9$ and that the maximum value at $r=0$ is now $\lambda_\mathrm{max}=\ln 3$.
If one computes the homeomorphism conjugating two maps of the family by a trivial
extension of the method (\ref{eq:succesive1})--(\ref{eq:succesive2}), one obtains a graph similar
to that of Fig.\ \ref{fig6}. The same happens if one uses   family (\ref{eq:family3}), 
in which case the Lyapunov coefficient is similar to that of Fig.\ \ref{fig4},
with the bifurcation at $r=1/9$ and the maximum value at $r=1$ given again by $\lambda_\mathrm{max}=\ln 3$.

For other values of the maximum Lyapunov exponent one can use a similar
method starting from a piecewise linear map with the desired value of the Lyapunov exponent.
 On the other hand, 
if a constant Lyapunov exponent
is needed, we can use instead the method leading to (\ref{eq:familyc}) but starting from an appropriate
chaotic map, such as (\ref{eq:fpiece}).

%%%%%%%%%%%%%%%%%%%%%%%%%%%%%%%%%%%%%%%%%%%%%%
\acknowledgments
%%%%%%%%%%%%%%%%%%%%%%%%%%%%%%%%%%%%%%%%%%%%%%
This work was supported by The University of the Basque Country
(Research Grant~GIU06/37).

\end{document}